\documentclass[
 aps,
 prl,
 twocolumn
]{revtex4-1}

\usepackage{graphicx}% Include figure files
\usepackage{dcolumn}% Align table columns on decimal point
\usepackage{bm}% bold math
\usepackage{hyperref}% add hypertext capabilities
\usepackage{color}
\usepackage[caption=false]{subfig}
\usepackage{microtype}
\usepackage{verbatim} 
\usepackage{amsmath,amsfonts,amssymb,amsthm}
\usepackage[algoruled, vlined, lined ,boxed,commentsnumbered]{algorithm2e}

\usepackage{tikz}
\hypersetup{%
        colorlinks=true,
        linkcolor=blue,
        citecolor=blue,
        urlcolor=blue,
      }
\usepackage[euler]{textgreek}
\usepackage[ugly]{units}
\usepackage{xcolor}
\usepackage{ wasysym }

\newcommand{\ket}[1]{|#1\rangle}
\newcommand{\bra}[1]{\langle#1|}

\usepackage{fancyhdr}
\theoremstyle{definition}
\begin{abstract}
Machine learning techniques have led to broad adoption of a statistical model of computing. The statistical distributions natively available on quantum processors are a superset of those available classically. Harnessing this attribute has the potential to accelerate or otherwise improve machine learning relative to purely classical performance. A key challenge toward that goal is learning to hybridize classical computing resources and traditional learning techniques with the emerging capabilities of general purpose quantum processors. Here, we demonstrate such hybridization by training a 19-qubit gate model processor to solve a clustering problem, a foundational challenge in unsupervised learning. We use the quantum approximate optimization algorithm in conjunction with a gradient-free Bayesian optimization to train the quantum machine. This quantum/classical hybrid algorithm shows robustness to realistic noise, and we find evidence that classical optimization can be used to train around both coherent and incoherent imperfections.
\end{abstract}

\begin{document}

\title{Unsupervised Machine Learning on a Hybrid Quantum Computer}

\author{J. S. Otterbach}
\author{R. Manenti}
\author{N. Alidoust}
\author{A. Bestwick}
\author{M. Block}
\author{B. Bloom}
\author{S. Caldwell}
\author{N. Didier}
\author{E. Schuyler Fried}
\author{S. Hong}
\author{P. Karalekas}
\author{C. B. Osborn}
\author{A. Papageorge}
\author{E. C. Peterson}
\author{G. Prawiroatmodjo}
\author{N. Rubin}
\author{Colm A. Ryan}
\author{D. Scarabelli}
\author{M. Scheer}
\author{E. A. Sete}
\author{P. Sivarajah}
\author{Robert S. Smith}
\author{A. Staley}
\author{N. Tezak}
\author{W. J. Zeng}
\author{A. Hudson}
\author{Blake R. Johnson}
\author{M. Reagor}
\author{M. P. da Silva}
\author{C. Rigetti}
\affiliation{%
Rigetti Computing, Inc., Berkeley, CA
}%

\date{\today}% It is always \today, today,

\maketitle

%\tableofcontents

\section*{Introduction}

The immense power of quantum computation is illustrated by flagship quantum algorithms that solve problems such as factoring~\cite{Shor1994} and linear systems of equations~\cite{Harrow2009}, amongst many others, much more efficiently than classical computers.
The building of a quantum device with error rates well below the fault-tolerance threshold~\cite{Shor1996,Knill1998,Aharonov2008,Aliferis2006} poses a challenge to the implementation of these kinds of quantum algorithms on near-term devices.
In recent years several new algorithms targeting these near-term devices have been proposed. These algorithms focus on short-depth parameterized quantum circuits, and use quantum computation as a subroutine embedded in a larger classical optimization loop. It has been shown that optimizing the performance of the quantum subroutine---by varying a few free parameters---allows for calculating binding energies in quantum chemistry~\cite{Peruzzo2014,McClean2016,Kandala2017,Colless2017}, as well as solving some combinatorial~\cite{Farhi2014,nasa-qaoa} and tensor network problems~\cite{Kim2017}.
In this paper we choose to focus on an unsupervised machine learning task known as {\em clustering}, which we translate into a combinatorial optimization problem~\cite{Lucas2014,Poljak1995} that can be solved by the quantum approximate optimization algorithm (QAOA)~\cite{Farhi2014,nasa-qaoa}. We implement said 
algorithm on a 19-qubit computer using a flexible quantum programming platform~\cite{Smith2016,pyQuil}. 
We show that our implementation of this algorithm finds the optimal solution to random 
problem instances with high probability, 
and that good approximate solutions are found in all investigated cases, even with
relatively noisy gates. This robustness is enabled partially by using a Bayesian procedure~\cite{Rasmussen2006,BayesOpt} 
in the classical optimization loop for the quantum circuit parameters.

\section*{Clustering}

The particular unsupervised machine learning problem we focus on here is known as clustering~\cite{Jain1999, Jain1988}. Clustering consists of assigning labels to elements of a dataset based only on how similar they are to each other---like objects will have the same label, unlike objects will have different labels. Mathematically, the dataset $\mathcal{D}$ has elements $\mathbf{x}_i$ for $1\le i \le n$, where each element is a $k$-dimensional {\em feature vector} (a numerical representation of any object of interest: a photograph, an audio recording, etc.).
In order to represent dissimilarity, we need to define a distance measure $d(\mathbf{x}_i,\mathbf{x}_j)$ between two samples $\mathbf{x}_i$ and $\mathbf{x}_j$. A familiar choice for a distance measure is the Euclidean distance, but specific applications may naturally lead to very different choices for the metric and the distance~\cite{Shirkhorshidi2015, Boriah2008}.

Most common choices of distances allows us to calculate a matrix of distances between all points in $\mathcal{O}(k n^2)$ steps, by simply calculating the distance between every possible pair of data samples. Let this matrix be denoted by $C$ where $C_{ij} = d(\mathbf{x}_i,\mathbf{x}_j)$. This matrix can be interpreted as an adjacency matrix of a graph $G$, where each vertex represents an element of $\mathcal{D}$ and $C_{ij}$ is the {\em weight} of edge between vertices $i$ and $j$. In general the matrix $C$ will be dense leading to a fully connected graph on $n$ vertices, but different choices of distance metrics along with coarse-graining can make this distance matrix sparse. In clustering, the main assumption is that distant points belong to different clusters; hence maximizing the overall sum of all weights (distances) between nodes with different labels represents a natural clustering algorithm. The mathematical formulation of this is a Maximum-Cut (\textsc{Maxcut}) problem~\cite{Poljak1995}, defined as $\textsc{Maxcut}(G, C)$ for the dense graph $G$ of the distance matrix $C$.

More precisely, the \textsc{Maxcut} problem consists of an undirected graph $G=(V,E)$ with a set of vertices $V$ and a set of edges $E$ between those vertices. The weight $w_{ij}$ of an edge between vertices $i$ and $j$ is a positive real number, with $w_{ij} = 0$ if there is no edge between them. A {\em cut} $\delta(S)\subset E$ is a set of edges that separates the vertices $V$ into two disjoint sets $S$ and $\bar{S}=V\setminus S$. The cost $w(\delta(S))$ of a cut is defined as the sum of all weights of edges connecting vertices in $S$ with vertices in $\bar{S}$
\begin{align}
    w(\delta(S)) = \sum_{i\in S, j\in \bar S} w_{ij}. \label{eq:cut_cost}
\end{align}
The problem $\textsc{Maxcut}(G,w)$ is now easily formulated as an optimization objective
\begin{align}
    \textsc{Maxcut}(G, w) = \max_{S\subset V} w(\delta(S)).
\end{align}

The \textsc{Maxcut} problem is an example of the class of NP-complete problems~\cite{Karp72}, which are notoriously hard to solve. Many other combinatorial problems can be reduced to \textsc{Maxcut}---e.g., machine scheduling~\cite{Bahram1994}, computer-aided design~\cite{Krarup1978}, traffic message management problems~\cite{Gallo1980}, image recognition~\cite{Neven2008}, quadratic unconstrained optimization problems (QUBO)~\cite{Poljak1995} and many more. One approach to solving \textsc{Maxcut} is to construct a physical system---typically a set of interacting spin-$\frac{1}{2}$ particles~\cite{Lucas2014}---whose lowest energy state encodes the solution to the problem, so that solving the problem is equivalent to finding the ground state of the system~\footnote{Note that we can turn any maximization procedure into a minimization procedure by simply changing the sign of the objective function.}. This is the approach we take here.

\section*{Quantum Approximate Optimization Algorithm}

\begin{figure}[t]
    \includegraphics[width=1\columnwidth]{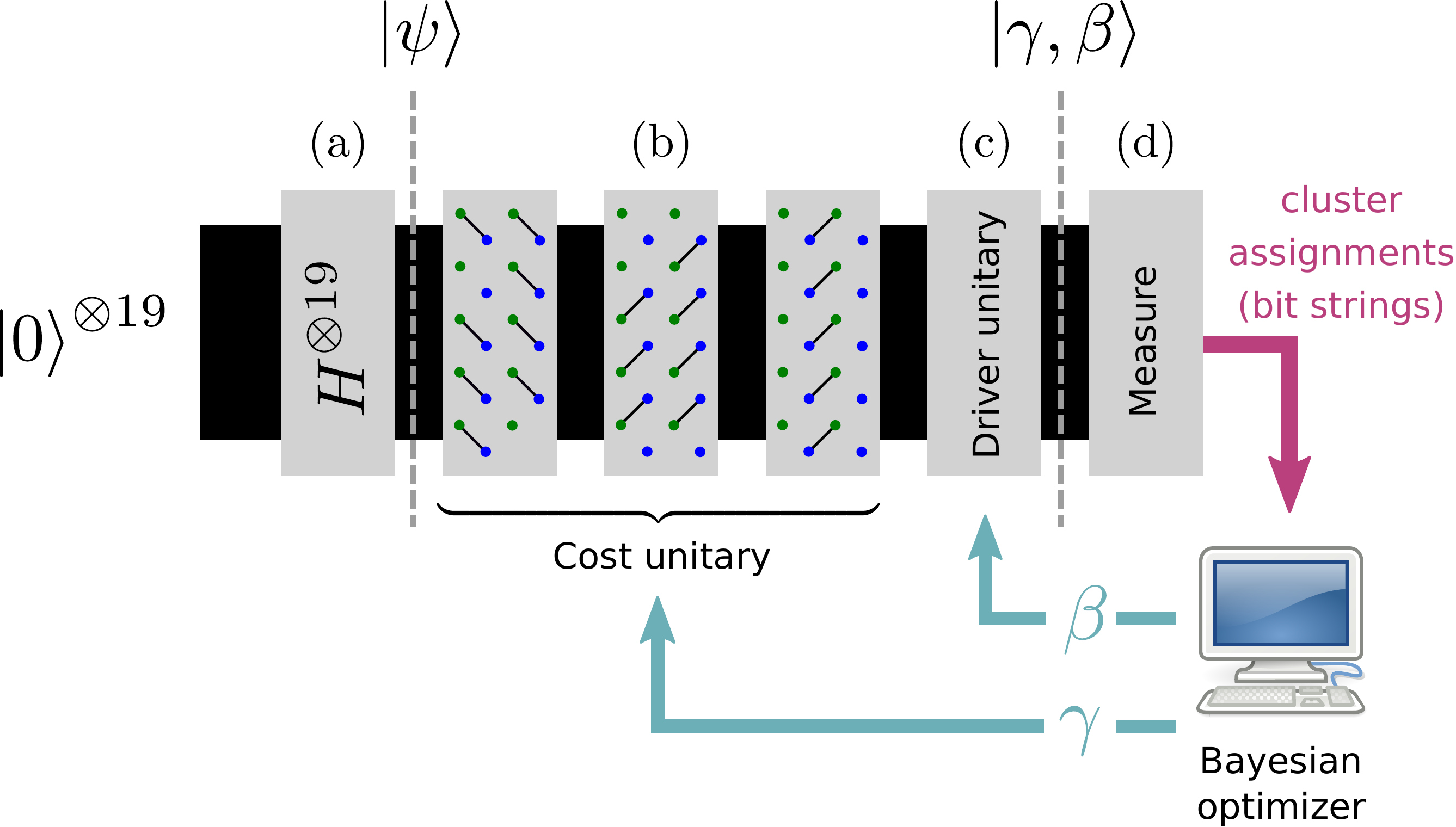}
    \caption{Circuit diagram for the clustering experiment described in the main text~\footnote{Computer icon is used under a Creative Commons license~\cite{gnome-icon}.}, which corresponds to a single iteration of 
    QAOA to solve a \textsc{Maxcut} problem. The algorithm proceeds by (a) initializing all qubits in the 0 state, and then 
    applying Hadamard gates to each of the qubits. (b) Next, due to parallelism constraints, it applies three independent rounds
    of interactions of the form $\exp(-i\gamma w_{ij}\hat{\sigma}^z_i\hat{\sigma}^z_j)$ (actual gate sequence not shown for brevity). The 
    overall scaling $\gamma$ for the 
    interactions is chosen and varied by the optimization routine. (c) Single qubit $\hat\sigma_X$ rotations are applied to each qubit, 
    again with an overall rotation angle $\beta$ chosen by the optimization routine. (d) Finally, all qubits are measured and fed 
    into the classical optimization routine. Many shots are collected, and the best \textsc{Maxcut} cost, along with the 
    history of previous angles, is used to inform the next choice for $(\beta, \gamma)$.}
    \label{fig:QAOA_circuit_ring_of_disagrees}
\end{figure}

\begin{figure}[t]
\includegraphics[width=0.95\columnwidth]{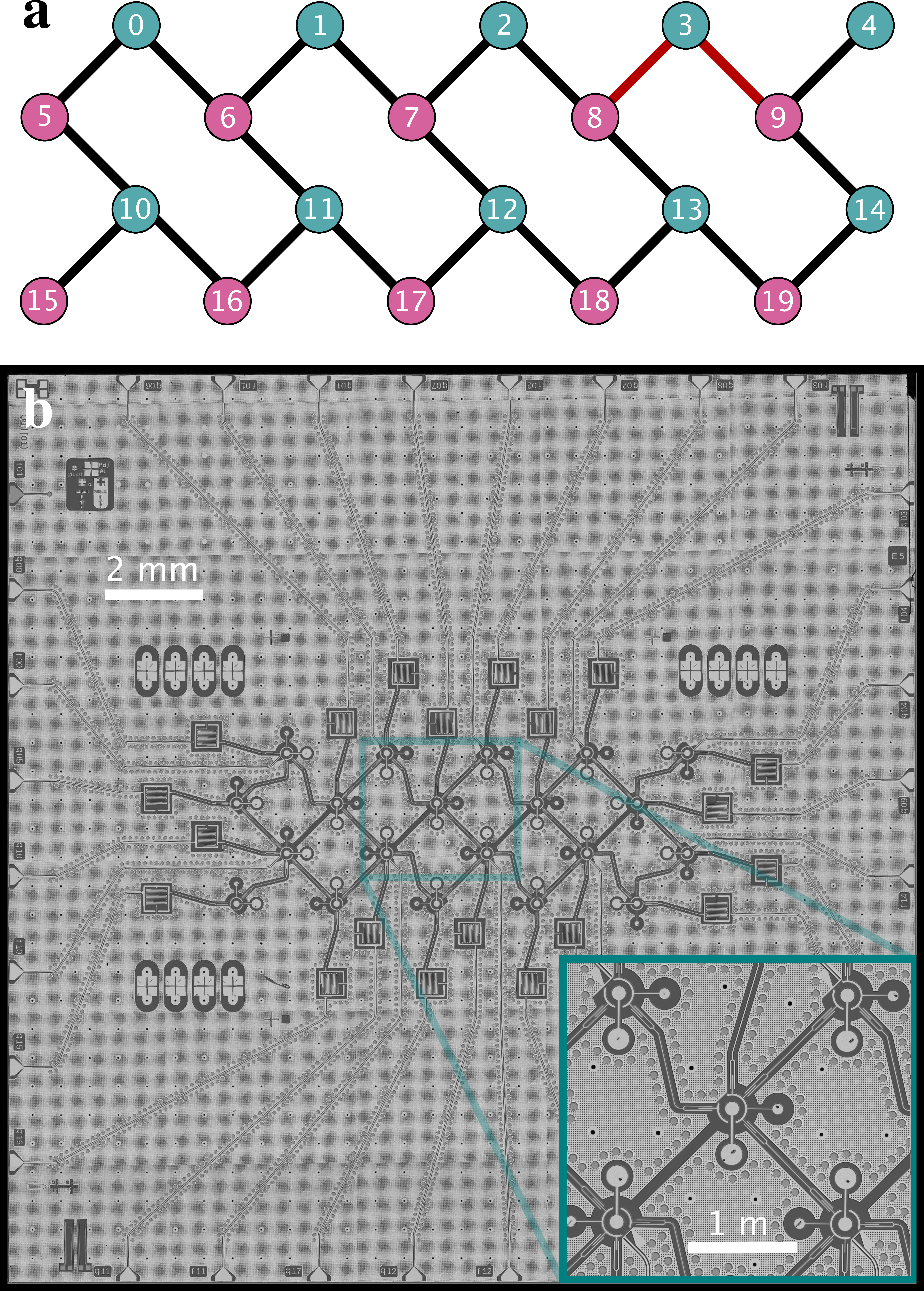}
\caption{
\label{fig:chip}
{\bf Connectivity of Rigetti 19Q}. {\bf a}, Chip schematic showing tunable
transmons (teal circles) capacitively coupled to fixed-frequency transmons
(pink circles). {\bf b}, Optical chip image. Note that some couplers have
been dropped to produce a lattice with three-fold, rather than four-fold,
connectivity.
}
\end{figure}

It is possible to find the ground state of interacting spin systems using an algorithm known as the {\em quantum approximate optimization algorithm} (QAOA)~\cite{Farhi2014}. QAOA can be thought of as a heuristic to prepare a superposition of bit strings with probability amplitudes heavily concentrated around the solution. The encoding of the problem itself is given by a {\em cost Hamiltonian} (cf.~\ref{eq:weighted_maxcut})
\begin{equation}
    \hat{H}_\textrm{C} = -\frac{1}{2}\sum_{i,j} C_{ij}(1-\hat{\sigma}_i^z\hat{\sigma}_j^z),
\end{equation}
and QAOA approximates the ground state by initially preparing the equal superposition of all bit strings, then iteratively applying a pair of unitary operations before measuring (see Fig.~\ref{fig:QAOA_circuit_ring_of_disagrees}). For the $i$th iteration, we evolve the system with cost unitary $\hat{U}_i=\exp(-i \gamma_i \hat{H}_\textrm{C} )$ for some angle $\gamma_i$, followed by the driver unitary $\hat{V}_i=\exp(-i \beta_i \hat{H}_\textrm{D} )$ for some angle $\beta_i$, where the {\em driver Hamiltonian} is
\begin{equation}
    \hat{H}_\textrm{D} = \sum_{i} \hat{\sigma}_i^x.
\end{equation}
In other words, QAOA prepares the state
\begin{equation}
    \ket{\pmb{\gamma},{\pmb \beta}} = \hat{V}_p \hat{U}_p \cdots \hat{V}_1 \hat{U}_1 \ket{\psi},
\end{equation}
where $p$ is the number of iterations, $\ket{\psi}$ is the uniform superposition of all bit strings (with the same phase), and measures this state in the computational basis.

For the optimal choice of angles, the probability of sampling the bit string corresponding to the lowest energy state of $\hat{H}_\textrm{C}$ increases monotonically with the number of iterations applied~\cite{Farhi2014}.  As the number of iterations approaches infinity, the probability of success approaches unity. However, QAOA does not specify how the optimal set of angles are found, although in some cases they can be derived analytically for a small number of iterations and for uniformly weighted graphs~\cite{Farhi2014,Hadfield2017}.

\section*{Bayesian optimization}

Our strategy to find the (approximately) optimal bit-string for a weighted \textsc{Maxcut} instance, corresponding to the solution of the clustering problem, uses Bayesian optimization over the QAOA angles $({\pmb{\gamma}}, {\pmb{\beta}})$~\cite{Shahriari2015,Snoek2015}. This framework constructs a prior distribution over functions, sequentially evaluates a black box objective---the \textsc{Maxcut} instance in this case---and updates the prior according to Bayes' rule
\begin{align}
    p(f|y) \sim p(y|f)p(f),
\end{align}
where $p(f|y)$ is the posterior distribution over function space given the observations $y$, $p(f)$ is the prior over function space, and $p(y|f)$ is the likelihood of observing the values $y$ given the model for $f$. With growing number of optimization steps (observations $y$) the true black-box objective is increasingly well approximated. The trick lies in choosing the prior $p(f)$ in a way that offers closed-form solutions for easy numerical updates, such as Gaussian processes, which assume a normal distribution as a prior over the function space~\cite{Rasmussen2006}(cf.~\ref{suppl:eq:gaussian_process}). In the present case of QAOA, it should be noted that sampling at each step will generally lead to a non-trivial distribution of values when the state $\ket{\pmb{\gamma},{\pmb \beta}}$ is entangled or mixed. To fit this into the Bayesian Optimization framework we calculate the best observed sample and return this to the optimizer. Hence, the function $f$ represents the value of the best sampled bit string at location $\pmb{\gamma},{\pmb \beta}$. More generally, one could compute any statistic of the distribution (as detailed in the appendix).

To avoid a random walk over the space of potential evaluation points, the Bayesian optimizer maximizes a utility function that can be calculated from the posterior distribution after each update. In this way, it intelligently chooses points to minimize the number of costly evaluations of the black box objective function (see the appendix for more details).

\section*{The quantum processor}

We ran the QAOA optimizer on a quantum processor consisting of 20 superconducting transmon qubits~\cite{Koch2007}
with fixed capacitive coupling in the lattice shown in Fig.~\ref{fig:chip}. Qubits 0--4 and
10--14 are tunable while qubits 5--9 and 15--19 are fixed-frequency devices. The
former have two Josephson junctions in an asymmetric SQUID geometry to provide
roughly $\unit[1]{GHz}$ of frequency tunability, and flux-insensitive ``sweet
spots''~\cite{Vion2002} near $\omega^{\textrm{max}}_{01}/2\pi \approx
\unit[4.5]{GHz}$ and $\omega^{\textrm{min}}_{01}/2\pi \approx \unit[3.0]{GHz}$.
These tunable qubits are coupled to bias lines for AC and DC flux delivery.
Each qubit is capacitively coupled to a quasi-lumped element resonator for
dispersive readout of the qubit state~\cite{Blais2004, Blais2007}. Single-qubit
control is effected by applying microwave drives at the resonator ports, and
two-qubit gates are activated via RF drives on the flux bias lines, described
below.

The device is fabricated on a high-resistivity silicon substrate with
superconducting through-silicon via technology~\cite{Vahidpour2017} to
improve RF-isolation. The superconducting circuitry is realized with Aluminum
($T_\textrm{c}\approx\unit[1.2]{K}$), and patterned using a combination of
optical and electron-beam lithography. Due to a fabrication defect, qubit 3 is
not tunable, which prohibits operation of the 2-qubit parametric gate described
below between qubits 3 and its neighbors (8 and 9). Consequently, we treat this
as a 19-qubit processor.

In Rigetti 19Q, as we call our device, each tunable qubit is capacitively coupled to one-to-three
fixed-frequency qubits. The DC flux biases are set close to zero flux such that
each tunable qubit is at its maximum frequency
$\omega^{\textrm{max}}_\textrm{T}$. Two-qubit parametric CZ gates are activated
in the $\ket{11} \leftrightarrow \ket{20}$ and/or $\ket{11} \leftrightarrow
\ket{02}$ sub-manifolds by applying an RF flux pulse with amplitude $A_0$,
frequency $\omega_\textrm{m}$ and duration $t_\textrm{CZ}$ to the tunable
qubit~\cite{Didier2017, Caldwell2017, Reagor2017}. For RF flux modulation about
the qubit extremal frequency, the oscillation frequency is doubled to
$2\omega_\textrm{m}$ and the mean effective qubit frequency shifts to
$\bar{\omega}_\textrm{T}$. Note that the frequency shift increases with larger
flux pulse amplitude. The effective detuning between neighboring qubits becomes
$\Delta = \bar{\omega}_\textrm{T} - \omega_\textrm{F}$.

The resonant condition for
a CZ gate is achieved when $\Delta = 2 \omega_\textrm{m} - \eta_\textrm{T}$ or
$\Delta = 2 \omega_\textrm{m} + \eta_\textrm{F}$, where $\eta_\textrm{T}$,
$\eta_\textrm{F}$ are the anharmonicities of the tunable and fixed qubit,
respectively. An effective rotation angle of $2\pi$ on these transitions imparts
a minus sign to the $\ket{11}$ state, implementing an effective CZ gate. The
time-scale of these entangling gates is in the range 100--250$\,\mathrm{ns}$. Due
to finite bandwidth constraints of our control electronics, the applied flux
pulse is shaped as a square pulse with linear rise and fall time of
$\unit[30]{ns}$.

\begin{figure}[t]
    \includegraphics[width=0.95\columnwidth]{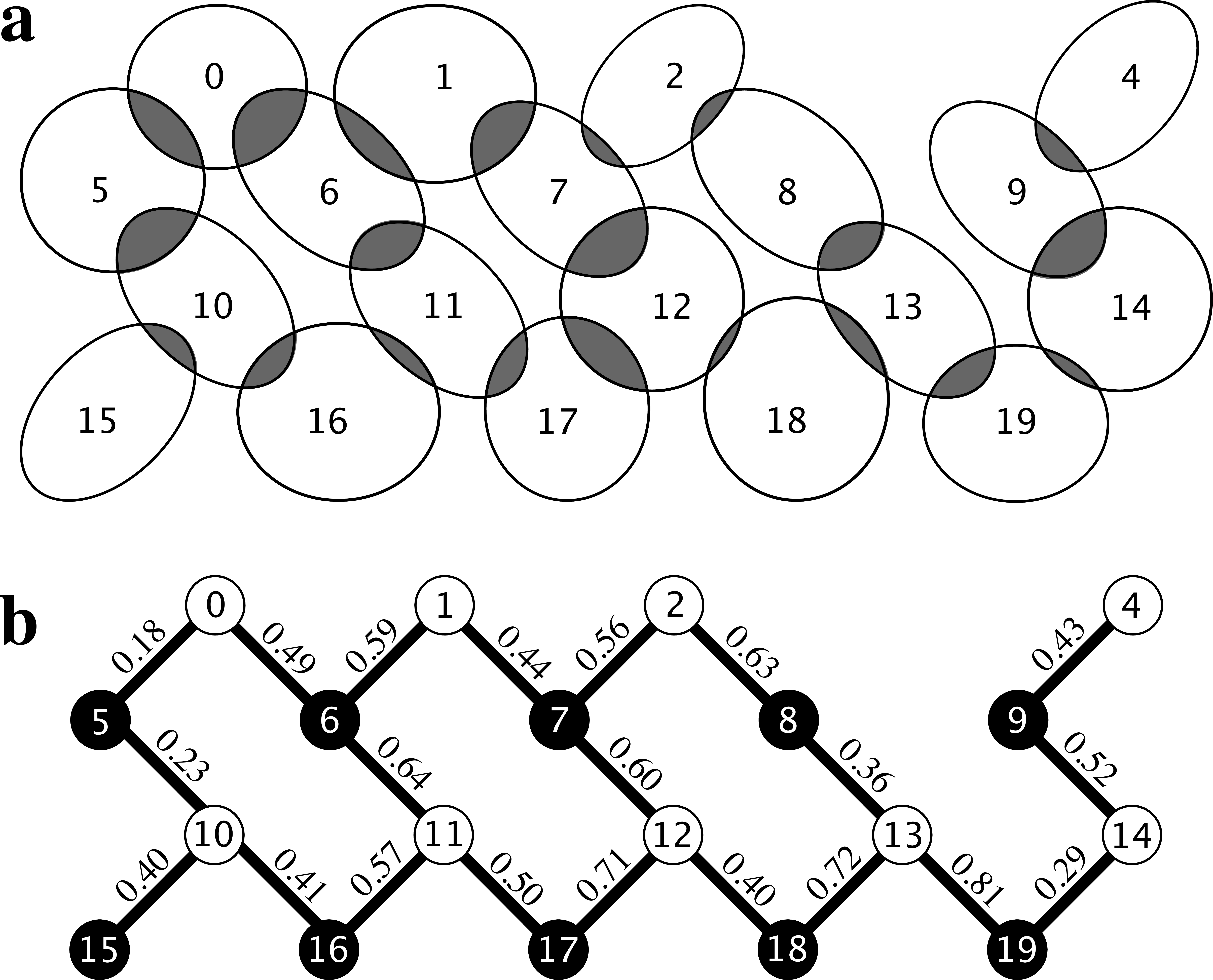}
    \caption{
    $\textbf{a},$ The general form of the clustering problem instance solved on the 19Q chip, and $\textbf{b},$ the corresponding \textsc{Maxcut} problem instance solved on the 19 qubit chip. The edge weights---corresponding to the overlap between neighbouring probability distributions on the plane---are chosen at random and the labels indicate the mapping to the corresponding qubit. The vertices are colored according to the solution to the problem.\label{fig:max-cut-instance}}
\end{figure}

\begin{figure}[t]
    \includegraphics[width=1\columnwidth]{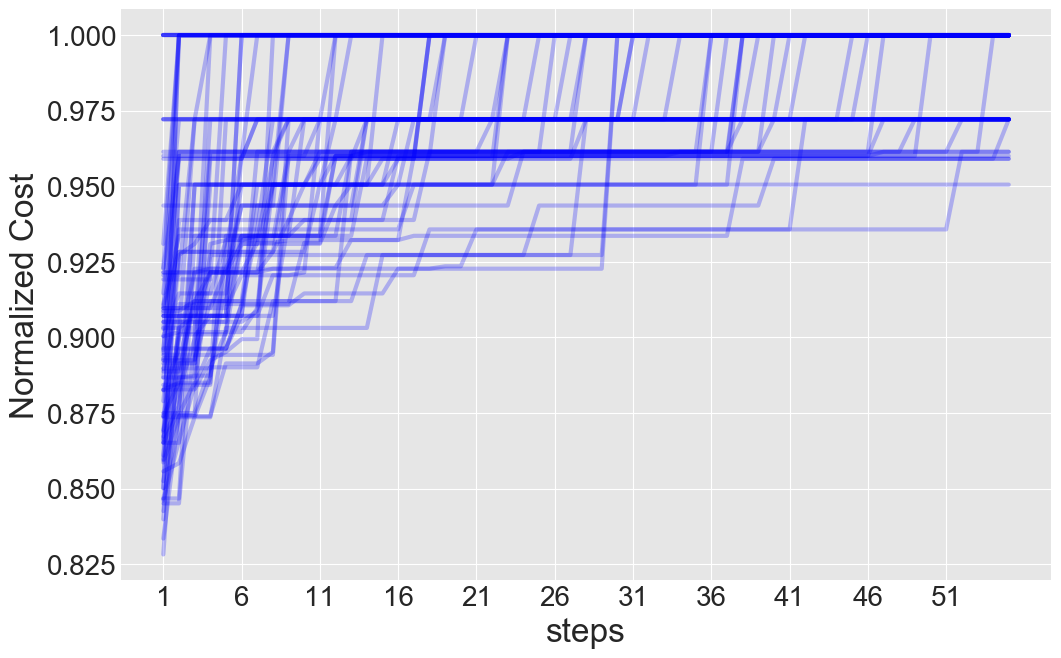}
    \caption{Traces for the normalized \textsc{Maxcut} cost for 83 independent runs of the algorithm on the 19Q chip for the fixed, but random, problem instances of Fig.~\ref{fig:max-cut-instance}. Notice that most traces reach the optimal value of $1$ well before the cutoff at 55 steps.}
    \label{fig:traces}
\end{figure}%

\begin{figure}[t]
    \includegraphics[width=1\columnwidth]{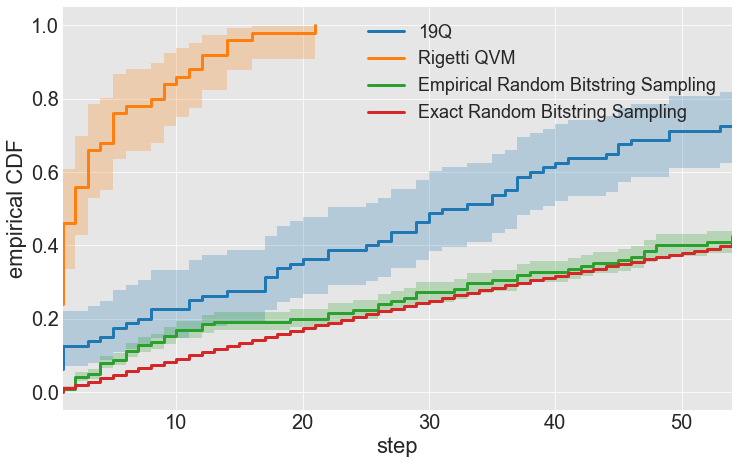}
    \caption{The performance of our implementation of the clustering algorithm on the 19Q chip (blue) and a noiseless simulation through the Forest~\cite{Forest} quantum virtual machine (orange) can be compared to the performance of an algorithm that simply draws cluster assignments at random (red: Theoretical curve, green: empirical confirmation). It is clear that out algorithm generates the optimal assignment much more quickly than it would be expected by chance: the $95\%$ confidence region for our empirical observations have very small overlap for the distribution given by random assignments. See appendix for more detailed statistics. \label{fig:ecdf-comparisson}}
\end{figure}

\section{Implementation}

We demonstrate the implementation of the proposed clustering algorithm on a problem with a cost Hamiltonian constructed to match the connectivity of Rigetti 19Q. Such a choice minimizes the circuit depth required for the algorithm while utilizing all available qubits. This problem corresponds to clustering an arrangement of overlapping probability distributions, whose dissimilarity is given by the Bhattacharyya coefficient~\cite{Bhattacharyya1943}---a measure for distributional overlap. A cartoon of such an arrangement, and the corresponding graph for the weighted \textsc{Maxcut} problem, is depicted in Fig.~\ref{fig:max-cut-instance}~\footnote{As the Bhattacharyya coefficient is not a traditional distance metric (it violates the triangle inequality) we should interpret clustering as the characteristic that distributions within the same cluster, i.e. the same label, have minimal---in our case zero---overlap. Phrased in this way the connection to VLSI design becomes obvious, where on sub-goal is to identify groups of objects with minimal overlap.}. This problem is solved using a single iteration ($p=1$) of QAOA, and using up to 55 steps of a Bayesian optimizer to choose the angles $(\gamma, \beta)$. This procedure is repeated several times to gather statistics about the number of steps necessary to reach the optimal answer.

Using the available interactions and local gate operations, we are able to implement the cost unitary in a circuit of depth corresponding to six \texttt{CZ} gates interspersed with single qubit operations and hence fitting well within a single coherence time of the qubits.  This circuit depth is dictated by two independent factors. The first is the implementation of the cost unitary. Since all the terms in the cost Hamiltonian commute, it is possible to decompose the cost unitary into a separate unitary for each cost term. These cost terms, in turn, take the form
$\exp(-i\gamma w_{ij}\hat{\sigma}^z_i\hat{\sigma}^z_j)$ that does not directly correspond to one of the native gates in our quantum computer. However, they can be
implemented by the sequence requiring two interaction $\texttt{CNOT}_{ij} \cdot \texttt{RZ}_j(\gamma w_{ij})\cdot \texttt{CNOT}_{ij}$~\cite{Nielsen2002}. The second factor
is that the $\texttt{CNOT}$ interactions cannot be applied simultaneously on overlapping sets of qubits. Given the connectivity of the graph
for our problem (where some vertices have 3 neighbors), that means the cost terms must be broken down into 3 rounds of non-overlapping unitaries, 
each of which consists of two interactions, so that the overall circuit has a depth corresponding to 6 two-qubit gates interspersed with single qubit operations. Additional circuit compilation steps are taken to minimize the number of intermediate single-qubit operations, so the depth is ultimately dominated by the interaction.

The entire algorithm is implemented in Python, leveraging the \texttt{pyQuil} library~\cite{pyQuil} for describing parameterized quantum circuits in the quantum instruction language \texttt{Quil}~\cite{Smith2016}, and run through the Forest platform~\cite{Forest} for controlling the quantum computer and accessing the data it generates. The Bayesian optimizer is provided by the open source package \texttt{BayesianOptimization}, also written in Python~\cite{BayesOpt}.

\section*{Results}

The typical behaviour of the algorithm is illustrated in Fig.~\ref{fig:traces}.
Each trace corresponds to a separate run of the algorithm on a fixed, but random, choice for the graph weights. In particular, it corresponds to the best observed cost up to the current Bayesian optimization step, i.e. the historic best. The costs are normalized such that $1$ corresponds to the optimal solution, which can be easily computed for the problem instance under consideration. 

The run time for 55 Bayesian optimization steps with $N=2500$ measurements per step is approximately 10 minutes. This run time includes network latency in the communication between the quantum processor and the classical processor running the high-level hybrid algorithm, as well as many other factors such as circuit compilation and job scheduling. For the random problem instances chosen, the vast majority of the traces reach the optimum in fewer than 55 steps (only about ~$25\%$ of runs did not reach the optimum within 55 steps, as illustrated in Fig.~\ref{fig:ecdf-comparisson}). Different random choices for the graph weights yield results that are qualitatively identical (see appendix for more details).

Quantitatively, we can compare how execution of the algorithm on the quantum computer compares with drawing a comparable number of bit strings uniformly at random, and remembering the ones with the highest cost. Random sampling of bit strings can be seen as a proxy for randomly walking the parameter space, which is expected to be a better strategy than grid-search~\cite{Bergstra2012}. In that case, the probability of obtaining the optimal cost should decrease exponentially with the number of points in the clustering problem, but increase monotonically with the total number of samples. 
In Fig.~\ref{fig:ecdf-comparisson} we compare the empirical distributions for a noiseless simulation of the algorithm (orange), the experimental implementation of the algorithm (blue), and the random sampling of bit strings with finite statistics (green), along with the analytical distribution for random sampling of bit strings (red)---the shaded regions correspond to $95\%$ confidence intervals. Taking the null hypothesis to be the random sampling of bit strings, standard hypothesis testing methods based on the Kolmogorov-Smirnov statistic exclude the null hypothesis as an explanation for the behavior of the experimental implementation of the algorithm to a level higher than $99\%$. Similarly, the null hypothesis is excluded as an explanation for the behaviour of the noiseless simulation of the algorithm to a level higher than $99.99\%$ (see the appendix for more details).

\section*{Conclusion and summary}

We have used a 19-qubit quantum computer to demonstrate a hybrid quantum algorithm for clustering. This algorithm is based on the quantum approximate optimization algorithm, which in turn can be applied to a broad class of combinatorial optimization problems, ranging from image recognition to machine scheduling. Our implementation relies on Bayesian optimization of classical parameters within the quantum circuit, and we show that the algorithm reaches the optimal solution in many fewer steps than would otherwise be expected by drawing cluster assignments uniformly at random. This demonstration uses the largest number of qubits to-date for an algorithm in a gate-based quantum processor. Harder combinatorial problems (larger instances and more complex topologies) may be handled by larger number of iterations of the quantum approximate optimization algorithm, which would necessitate both higher gate fidelities and higher qubit counts. Given the complexity of solving combinatorial problems, relatively modest improvements on these fronts could lead to the demonstration of a hybrid algorithm that outperforms classical algorithms for the same problem.

\section*{Acknowledgements}

The algorithm was written in Python with \texttt{pyQuil}~\cite{pyQuil} and uses the \texttt{BayesianOptimization} package~\cite{BayesOpt}. The data analysis was performed using code written in \texttt{NumPy}~\cite{numpy}, \texttt{QuTiP}~\cite{qutip}, and the figures were made with \texttt{matplotlib}~\cite{matplotlib} and Inkscape~\cite{inkscape}.

\section*{Contributions}

J.S.O., N.R., and M.P.S. developed the theoretical proposal. J.S.O. and E.S.F. implemented the algorithm, and J.S.O. performed the data analysis. M.B., E.A.S., M.S., and A.B. designed the 20-qubit device. R.M., S.C., C.A.R., N.A., A.S., S.H., N.D., D.S., and P.S. brought up the experiment. C.B.O., A.P., B.B., P.K., G.P., N.T., M.R. developed the infrastructure for automatic re-calibration. E.C.P., P.K., W.Z., and R.S.S. developed the compiler and QVM tools. 
J.S.O., M.P.S., B.R.J., M.R., R.M. wrote the manuscript. B.R.J., M.R., A.H., M.P.S., and C.R. were principal investigators of the effort.

%\bibliography{qluster_bib}


\clearpage

%%%%%%%%%%
%%%%%%%%%%
%%%%%%%%%%
%%%%%%%%%%
%%%%%%%%%%

%%%%%%%%%% Merge with supplemental materials %%%%%%%%%%
\pagebreak
\widetext
\begin{center}
\section{\large Supplementary Information}
\end{center}
%%%%%%%%%% Merge with supplemental materials %%%%%%%%%%
%%%%%%%%%% Prefix a "S" to all equations, figures, tables and reset the counter %%%%%%%%%%
\setcounter{equation}{0}
\setcounter{figure}{0}
\setcounter{table}{0}
\setcounter{page}{1}
\makeatletter
\renewcommand{\theequation}{S\arabic{equation}}
\renewcommand{\thefigure}{S\arabic{figure}}
\renewcommand{\thetable}{S\arabic{table}}
\renewcommand{\bibnumfmt}[1]{[#1]}
\renewcommand{\citenumfont}[1]{#1}
%%%%%%%%%% Prefix a "S" to all equations, figures, tables and reset the counter %%%%%%%%%%

\subsection{Ising Hamiltionian of the Maxcut problem}\label{suppl:MaxcutIsingDerivation}

Starting with the \textsc{Maxcut} formulation (\ref{eq:cut_cost}) we can construct the Ising Hamiltonian connected to a given \textsc{Maxcut} instance. To this end we note that we can lift a general graph $G$ on $n$-nodes to a fully connected graph $K_n$ by introducing missing edges and initializing their corresponding weights to zero. We assume that the weights $w_{ij} = w_{ji}$ are symmetric, corresponding to an undirected graph and introduce Ising spin variables $s_j \in \{-1, +1\}$ taking on value $s_j = +1$ if $v_j \in S$ and $s_j = -1$ if $v_j \in \bar S$. With this we can express the cost of a cut as
\begin{align}
    w(\delta(S)) = & \sum_{i\in S, j\in \bar S} \! w_{ij} \nonumber\\
                 = & \frac{1}{2}\sum_{(i, j)\in \delta(S)} \! w_{ij} \nonumber \\
                 = & \frac{1}{4}\sum_{i, j\in V} w_{ij} - \frac{1}{4} \sum_{i, j\in V} w_{ij} s_i s_j\nonumber\\
                 = & \frac{1}{4}\sum_{i, j\in V} w_{ij} (1- s_i s_j). \label{eq:weighted_maxcut}
\end{align}\\
Identifying the spin variables with the spin operators of qubits yields the quantum analog of the weighted \textsc{Maxcut} problem as
\begin{align}
    \hat{H}_\textrm{C} = - \frac{1}{2} \sum_{i, j\in V} w_{ij} (1- \hat{\sigma}^z_i \hat{\sigma}^z_j) \label{eq:maxcut_hamiltonian}
\end{align}
where we introduce the additional ``$-$'' sign to encode the optimal solution as the minimal energy state (as opposed to the maximal energy state from the \textsc{Maxcut} prescription). In this sense we want to minimize $\hat{H}_\textrm{C}$.

\subsection{Detailed sampling procedure}

\begin{figure*}[t]
    \includegraphics[width=0.9\textwidth]{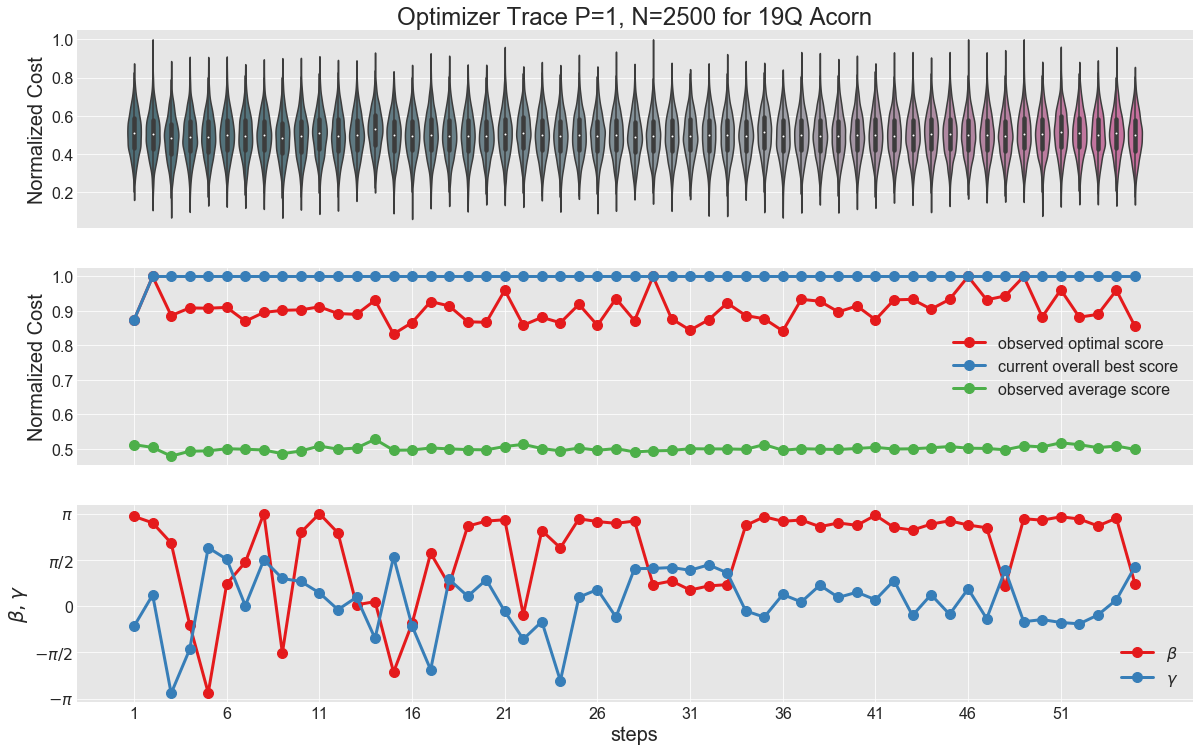}
    \caption{Trace of the Bayesian Optimization of a $p=1$ step QAOA procedure for the 19 qubit \textsc{Maxcut} problem instance with random weights as discussed in the main text. Each violin contains $2500$ samples drawn from the QPU and is cut off at its observed extreme values. We normalized the plot to indicate the best possible value. Note that this is intractable in general (i.e., it requires knowledge of the true optimum, which is hard to obtain). Detailed descriptions are in the text.}
    \label{app:fig:14Q_Qluster_QPU}
\end{figure*}

The full optimization trace of running the Bayesian optimized \textsc{Maxcut} clustering algorithm is shown in Fig.~\ref{app:fig:14Q_Qluster_QPU}, where the abscissa shows the step count of the optimizer. Each violin in the top panel shows the kernel-density estimates (KDE) of the cost distribution associated with the sampled bit-strings at the corresponding step. The width reflects the frequency with which a given cost has been sampled, while the thick and thin line within each violin indicate the $1\sigma$ and $2\sigma$ intervals of the distribution, respectively. To indicate the extreme values we cut off the KDE at the extreme values of the sample. Finally the white dot at the center of the violins show the mean value of the sampled distribution. In the optimization procedure we return the largest value of the cost distribution. The middle panel shows the best sampled value at step $i$ (red curve) corresponding to the extreme value of the distributions in the top panel, whereas the green curve is the mean value. The blue curve is the historic best value of the optimizer and shows the construction of the individual trace curves of Fig.~\ref{fig:traces}. Finally the lowest panel shows the behavior of the Bayesian optimizer in choosing the next hyperparameter pair $(\gamma, \beta)$. The jumpiness of the angle choices is likely due to the noise in the 19Q chip and seems significantly reduced in a simulated experiment as seen in Fig.~\ref{fig:20Q_FullyConnectedQlusterOptimizerTrace}.

As we can see, there is some variability in the mean value as well as the width of the distributions. At certain $(\beta, \gamma)$ points we do indeed sample large cost values from the distribution corresponding to (approximate) solutions of the randomly chosen \textsc{Maxcut} problem instance.

\begin{algorithm}[b]
\SetAlgoLined
\SetInd{0.5em}{0.5em}
\SetKwData{ch}{$H_\text{MC}$}
\SetKwData{distmat}{$C$}
\SetKwData{distmatij}{$C_{ij}$}
\SetKwData{bs}{bitString}
\SetKwData{clustS}{$\mathcal{D}_{S}$}
\SetKwData{clustNotS}{$\mathcal{D}_{\bar S}$}
\SetKwFunction{Encode}{Encode}
\SetKwFunction{MaxCutQAOA}{MaxCutQAOA}
\SetKwFunction{Distance}{Distance}
\KwData{Dataset $\mathcal{D}$ of points $p_i, i=1,\dots,N$}
\KwResult{bi-clustering assignments for the dataset $\mathcal{D}$ into $S$ and $\bar S$}
    \BlankLine
    \For{$i \leftarrow 1$ \KwTo $N$}{
        \For{$j \leftarrow 1$ \KwTo $N$}{
            \distmatij $\leftarrow$ \Distance{$p_i$, $p_j$}
        }
    }
    \BlankLine
    \ch $\leftarrow$ \Encode{\distmat} \;
    \bs $\leftarrow$ \MaxCutQAOA{\ch}\;
    \BlankLine
    \For{$i \leftarrow 1$ \KwTo $N$}{
        \eIf{\bs$[i] == 1$}{\clustS.append($p_i$)}{\clustNotS.append($p_0$)}
    }
    \Return \clustS, \clustNotS\;
\caption{\textsc{Maxcut} bi-clustering}\label{alg:OneStepQluster}
\end{algorithm}

\subsection{Clustering on a fully connected graph}

\begin{figure}[t!]
    \centering
    \subfloat[\label{fig:DistanceMatrixForQluster}]{
        \includegraphics[width=0.48\textwidth]{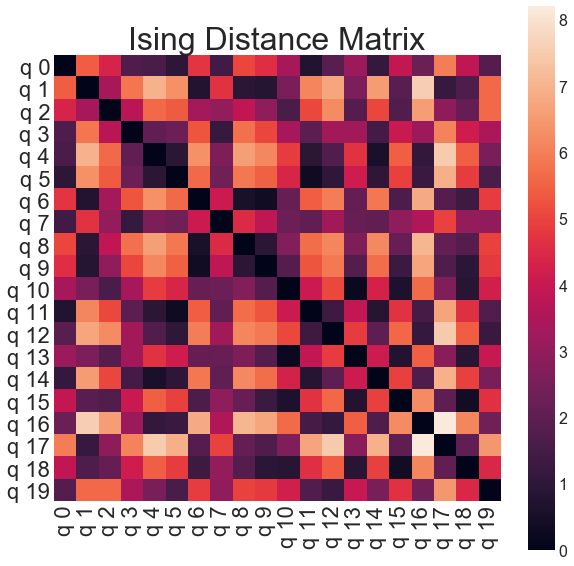}
    }
    ~
    \subfloat[\label{suppl:fig:QlusterLabelAssignment}]{
        \includegraphics[width=0.48\textwidth]{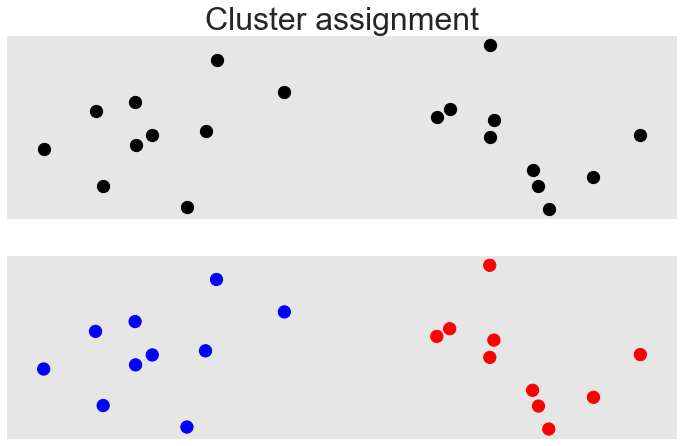}
    }

    \subfloat[\label{fig:20Q_FullyConnectedQlusterGraph}]{
        \includegraphics[width=0.48\textwidth]{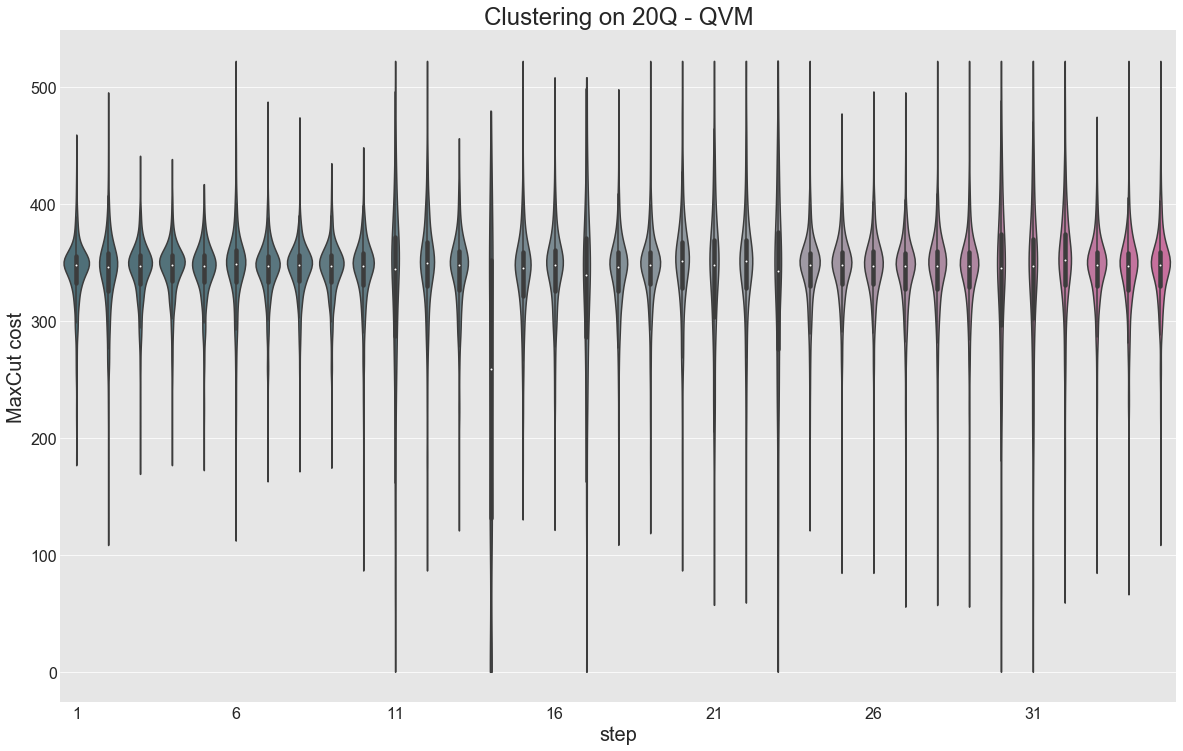}
    }
    ~
    \subfloat[\label{fig:20Q_FullyConnectedQlusterOptimizerTrace}]{
        \includegraphics[width=0.48\textwidth]{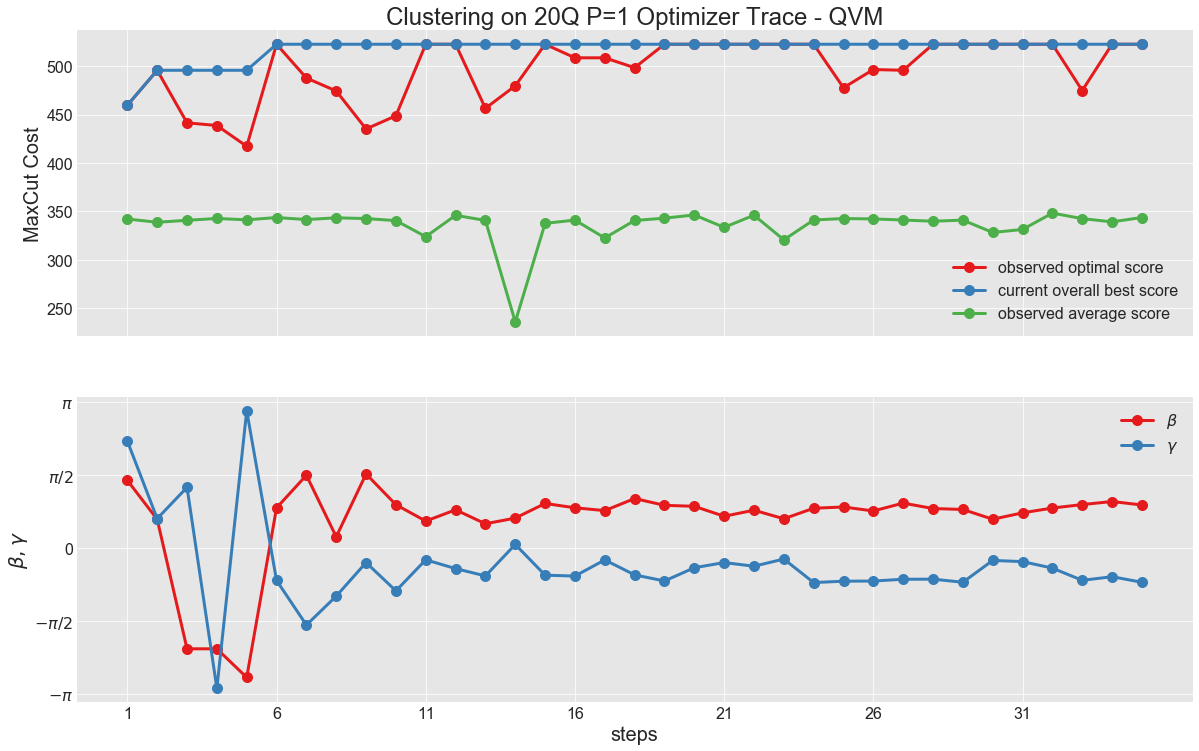}
    }
    \caption{
        (\textit{a}) Euclidean distance matrix for a sample of 20 points as shown in Fig.~\ref{suppl:fig:QlusterLabelAssignment} before labels have been assigned. This matrix will be used as the adjacency matrix for the \textsc{Maxcut} clustering algorithm.
        (\textit{b}) Random sample of 20 two-dimensional points forming two visually distinguishable clusters. The color assignment is the result of the clustering algorithm~\ref{alg:OneStepQluster}. Calculating the mutual Euclidean distances for all points gives rise to the distance matrix shown in Fig.~\ref{fig:DistanceMatrixForQluster}.
        (\textit{c}) QAOA optimization trace for a fully connected 20 node graph corresponding to the distance matrix in Fig.~\ref{fig:DistanceMatrixForQluster}. The parameters are $p=1$ and each violin contains $N=250$ samples, i.e. we sample $250 / 2^{20} \approx 0.02 $ of the whole state space. This demonstrates that the algorithm is capable of finding good solutions even for a non-trivial instance.
        (\textit{d}) QAOA optimization trace for a fully connected 20 node graph corresponding to the distance matrix in Fig.~\ref{fig:DistanceMatrixForQluster}. The parameters are $p=1$ and each violin contains $N=250$ samples. This demonstrates that the algorithm is capable of finding good solutions even for a non-trivial instance.
    }
\end{figure}

To demonstrate the clustering properties of the algorithm we simulate a larger \textsc{Maxcut} problem instance on the Quantum Virtual Machine (QVM). We construct the distance matrix shown in Fig.~\ref{fig:DistanceMatrixForQluster} resulting from the Euclidean distance between 20 random points in $\mathbb{R}^2$ as shown in Fig.~\ref{suppl:fig:QlusterLabelAssignment}. The corresponding graph is fully connected and the label assignment corresponding to its \textsc{Maxcut} solution is shown in Fig.~\ref{suppl:fig:QlusterLabelAssignment}. It is worth pointing out that this is a bipartite graph and has only two equally optimal solutions, the one shown and the exact opposite coloring. Hence randomly sampling bit-strings only has a chance of $2 / 2^{20} \approx 2 \cdot 10^{-6}$ of finding an optimal solution, meaning we would have to sample on the order of $2^{19}$ bit-strings to find the correct answer with significant success probability. The corresponding optimization trace is shown in Fig.~\ref{fig:20Q_FullyConnectedQlusterGraph}. Each violin contains $N=250$ samples, and hence we sample only $250 / 2^{20} \approx 0.02 $ of the full state space at each point corresponding to a chance $250\cdot2 / 2^{20} \approx 4.7\cdot 10^{-4}$ to sample the right bit-string. This corresponds to a $100\times$ improvement of the sampling procedure given a correctly prepared distribution as compared to just sampling from a uniform distribution. We can see that due to the fully connected nature of the graph the variation in the mean is not significant for $p=1$ steps in the QAOA iteration (see Fig.~\ref{fig:20Q_FullyConnectedQlusterOptimizerTrace} for more details). However, there is significant variations in the standard deviation of the sampled distributions with only a few of them allowing access to the optimal value with so few samples. A better view of the optimizer trace is shown in Fig.~\ref{fig:20Q_FullyConnectedQlusterOptimizerTrace} where we plot the average and best observed cost at each $(\beta, \gamma)$-pair in addition to the overall best value observed at the time a new point is evaluated. We can see that the optimizer slowly improves its best value and that it increasingly samples from distributions with large standard deviations. The clustering steps are described in pseudo-code by algorithm~\ref{alg:OneStepQluster}

\begin{figure*}[t!]
    \centering
    \subfloat[\label{fig:bayes_update_prior}]{
        \includegraphics[width=0.48\textwidth]{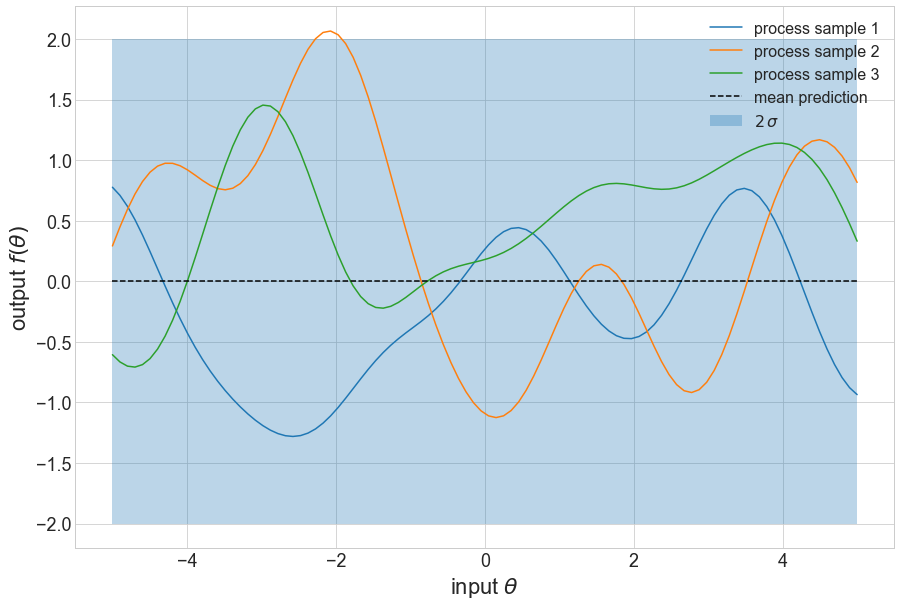}
        }
    ~
    \subfloat[\label{fig:bayes_update_posterior}]{
        \includegraphics[width=0.48\textwidth]{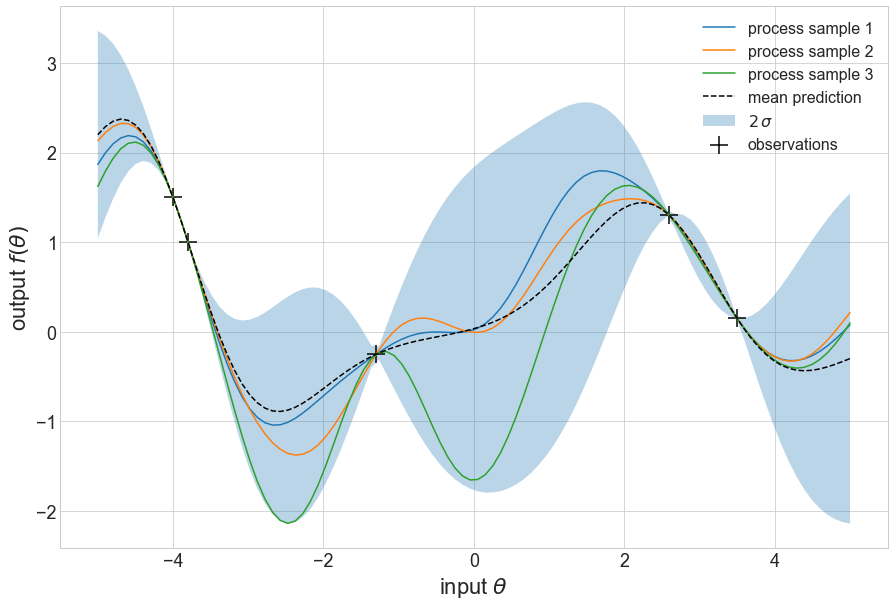}
    }
    \label{fig:bo_update}
    \caption{Update process of a Bayesian Optimization procedure. In both plots the blue area indicate a $2\sigma$ interval of the GP distribution. (\textit{a}) GP distribution prior; without any updates we can draw any of the three curves which more or less are within the two-standard deviation of the prior. (\textit{b}) Evaluating the true value of the random variable will \textit{clamp} the GP distribution at the observation points, forcing us to update the priors according to Bayes' rule. Drawing values from the posterior GP distribution will hence have a reduced overall uncertainty and adjusted mean value.}
\end{figure*}

\subsection{Construction of the optimization functional}

We can interpret the cost associated with the Ising Hamiltonian $\hat{H}_\textrm{C}$ corresponding to a classical combinatorial optimization problem (COP) as a function

\begin{align}
    f: \mathbb{Z}_2^n \times [0, 2\pi)^{2p} \rightarrow \mathbb{R}, f(x;\theta) \mapsto \bra{x}\hat{H}_\textrm{C}\ket{x},
\end{align}
where $\ket{x}\in\mathbb{Z}_2^n$ is a classic bit-string sampled from the distribution $\mathcal{D}_x$ with PDF $p(x;\theta) = |\psi(x;\theta)|^2$ prepared by running the QAOA procedure as described in the main text and $\theta = (\pmb\gamma,\pmb\beta)$. We can now identify the bit-string variable $\ket{x}$ as a random variable $X\sim \mathcal{D}_x(\theta)$ drawn from the parameterized distribution $\mathcal{D}_x(\theta) \subseteq \mathbb{Z}_2^n$ and consequently can interpret the function $f$ as a random variable $F\sim\mathcal{D}_f(\theta) \subseteq \mathbb{R}$ by functional composition with a PDF $p_f(v;\theta)$ and CDF $C_f(v;\theta)$

In order to optimize the distribution $\mathcal{D}_f(\theta)$ for sampling (nearly) optimal bit-strings we need to define an optimization objective. To this end, we consider the $j$-th order statistics $F_{(j)}$ \cite{Blitzstein} of an i.i.d. sample of $N$ experiments $\{F_i\}_{i=1\dots N}$
\begin{align}
    F_{(1)} = & \min \{F_1, \dots, F_N\} \nonumber \\
    F_{(2)} = & \min \{F_1, \dots, F_N\}\setminus \{F_{(1)}\} \nonumber \\
    \vdots&  \nonumber \\
    F_{(N-1)} = & \max \{F_1, \dots, F_N\}\setminus \{F_{(N)}\} \nonumber \\
    F_{(N)} = & \max \{F_1, \dots, F_N\}.
\end{align}
By definition we have $F_{(1)} \leq F_{(2)} \leq \dots \leq F_{(N)}$. Notice that the $F_{(j)}$ are random variables as well, but due to the ordering relation are not i.i.d. anymore. We can use the order statistic to define the $k$-th percentile ($k=1, ..., 100$) as the CDF of $F_{(\lfloor 100 j/N\rfloor = k)}$. The optimization routine can now be used to optimize the expectation values of any order statistic at hand. To this end we need to compute the PDF of these order statistics \cite{Blitzstein} according to 

\begin{align}
    p_{F_{(j)}} (v; \theta) = N & \binom{N-1}{j-1} p_f(v;\theta)  
      C_f(v;\theta)^{j-1} (1-C_f(v;\theta))^{N-j}
\end{align}

In practice we are mostly interested in optimizing the extreme value statistic, i.e. minimizing the first-order or maximizing the $N$-th order statistic. The expectation values of these can be computed as

\begin{align}
    s_1(\theta) := \langle F_{(1)}(\theta) \rangle = N \int\text{d}v\; v p_f(v;\theta)(1-C_f(v;\theta))^{N-1} \label{eq:MinValueStat}
\end{align}
and

\begin{align}
    s_N(\theta) :=\langle F_{(N)}(\theta) \rangle = N \int\text{d}v\; v p_f(v;\theta)C_f(v;\theta)^{N-1} \label{eq:MaxValueStat}
\end{align}

Note that this approach also enables us to estimate the uncertainty of these random variables, giving quality estimates of the sample. Despite looking horribly complicated, those values can readily be computed numerically from a set of samples of the distribution $\mathcal{D}_f(\theta)$. A pseudo-code representation of the statistics calculation is given in algorithm~\ref{alg:sample_extreme_val_statistic}.

\begin{algorithm}[b]
\SetAlgoLined
\SetInd{0.5em}{0.5em}
\KwData{QAOA angle parameters $\theta$}
\SetKwData{BS}{BitSamples}
\SetKwData{bs}{bitString}
\SetKwData{CV}{CostValues}
\SetKwData{cv}{cost}
\SetKwData{EL}{empty List}
\SetKwData{stat}{$s(\theta)$}
\SetKwFunction{Sample}{Sample}
\SetKwFunction{Cost}{Cost}
\SetKwFunction{Statistic}{Statistic}
\KwResult{value of a statistic $s(\theta)$ sampled from $\mathcal{D}_f(\theta)$}
\BlankLine
    \CV $\leftarrow$ \EL\;
    \For{$i \leftarrow 1$ \KwTo $N$}{
        \bs $\leftarrow$ \Sample{$\mathcal{D}_x(\theta)$}\;
        \cv $\leftarrow$ \Cost{\bs}\;
        \CV.append(\cv)\;
    }
    \tcp{calculate the statistic of interest}
    \stat $\leftarrow$ \Statistic{\CV}\; %\tcp*[r]{calculate the statistic of interest}
    \Return \stat\;
\caption{Statistics Calculation of Bit-String Distribution}\label{alg:sample_extreme_val_statistic}
\end{algorithm}

\subsection{Gaussian Process description of the extreme value optimization}\label{suppl:bayesian_optimization}

The extreme value functions $s_j(\theta): [0, 2\pi)^{2p} \rightarrow \mathbb{R}, j = 1, N$ are generally not analytically known and typically expensive to evaluate. Hence we have access to $s(\theta)$ only through evaluating it on a set of $m$ points $\theta_{1:m}$ \footnote{We use the short-hand notation $z_{i:j}$ to denote the collection of values $\{z_i,\dots z_j\}$} with corresponding variables $v_i = s(\theta_i)$ and noisy observations $y_{1:m}$. Note that we drop the subscript $j$ to denote the distinction between minimal and maximal value functions as the approach is identical for either of them. To make use of Bayesian optimization techniques \cite{Snoek2012, Shahriari2015} we assume that the variables $\mathbf{v}=v_{1:m}$ are jointly Gaussian and that the observations $y_{1:m}$ are normally distributed given $\mathbf{v}$, completing the construction of a Gaussian process (GP)\cite{Rasmussen2006}. The distinction between the variable $v_i$ and its observation $y_i$ is important given that the expectation values~\ref{eq:MinValueStat} and~\ref{eq:MaxValueStat} are subject to sampling noise, due to finite samples, but also due to finite gate and readout fidelities and other experimental realities, and hence cannot be known exactly. We describe the GP using the moments of a multivariate normal distribution
\begin{align}
    m(\theta) = & \mathbb{E}[v(\theta)]\\
    k(\theta, \theta') = & \mathbb{E}[(v(\theta)-\mu(\theta)) (v(\theta') - \mu(\theta'))]
\end{align}
and introduce the notation

\begin{align}
    v(\theta) \sim \mathcal{GP}(m(\theta), k(\theta, \theta'))
    \label{suppl:eq:gaussian_process}
\end{align}
This result summarizes the basic assumption that the extreme value statistic of \textit{the cost of best sampled bit-string} follows a Gaussian distribution as a function of the parameters $\theta$. While this might not be true, it proves to be a good choice for the unbiased prior in practice. To use the GP we need to specify the kernel function $k(\theta, \theta')$, which specifies a measure of correlation between two observation points $\theta$ and $\theta'$. Following \cite{Snoek2012} we use the Mat\'ern-2.5 kernel in the following, but there are many alternative choices \cite{Rasmussen2006, Shahriari2015}.

The basic idea of Bayesian Optimization is to draw new observations $y_*$ at points $\theta_*$ of the GP random variable $v$ and iteratively use Bayes' rule to condition the prior of $v$ on $y_*$. This conditioning is particularly simple for GPs as it can be performed analytically using the properties of multivariate normal distributions.

To account for imperfect/noisy evaluations of the true underlying function we simply need to adjust the update rule for the GP mean and kernels. This can also be done analytically and hence be directly applied to our case of numerically sampled extreme values statistics. For exhaustive details of the update rules, kernels and general GP properties see Ref.~\cite{Rasmussen2006}. It should be noted that the updated rules of Gaussian kernels require a matrix inversion which scales as $\mathcal{O}(m^3)$ and hence can become prohibitively expensive when the number $m$ of samples becomes large. Improving this scaling is an active area of research and early promising results such as Deep Networks for Global Optimization (DNGO) \cite{Snoek2015} provide surrogate methods for Gaussian processes with linear scaling. 

So far we have only been concerned with updating the GP to reflect the incorporation of new knowledge. To close the optimization loop we need a procedure to select the next sampling point $\theta_*$. Choosing a point at random will essentially induce a random walk over the optimization parameter domain and might not be very efficient \cite{Bergstra2012} (though this is still better than a grid search). A nice improvement over this random walk is offered by Bayesian framework itself: due to the constantly updating the posterior distribution we can estimate the areas of highest uncertainty and consequently chose the next point accordingly. However, this might bring the optimizer far away from the optimal point by trying to minimize the global uncertainty. To prevent the optimizer from drifting off we need a way to balance its tendency to explore areas of high uncertainty with exploiting the search around the currently best known value. This procedure is encapsulated in the acquisition function $\alpha(\theta; \mathcal{D}_m)$ where $\mathcal{D}_m = \{(\theta_i, y_i)\}_{i=1,\dots,m}$ is the set of observations up to iteration $m$ \cite{Shahriari2015}. Since the posterior is Gaussian for every point $\theta$ there are many analytic ways to construct an acquisition function. Here, we use the \textit{Upper Confidence Bound} (UCB) metric, which can be calculated as

\begin{align}
    \alpha_\text{UCB}(\theta; \mathcal{D}_m) = \mu_m(\theta) + \beta_m\sigma_m(\theta)
    \label{suppl:eq:acquisition_function}
\end{align}
where $\beta_m$ is a hyperparameter controlling the explore-exploit behavior of the optimizer and $\mu_m(\theta)$, $\sigma_m(\theta)$ are the mean and variance of the Gaussian of the posterior GP restricted to point $\theta$. Maximizing the acquisition function of all values $\theta$ at each iteration step yields the next point for sampling from the unknown function $s$. For more details see Ref.~\cite{Shahriari2015}. A pseudo-code representation of the Bayesian Optimization routine is given in algorithm~\ref{alg:bayes_opt}.

\begin{algorithm}[b]
\SetAlgoLined
\SetInd{0.5em}{0.5em}
\KwData{statistics function $f_s$ and parameter range $\mathcal{R}_\theta$}
\SetKwData{stat}{$s(\theta)$}
\SetKwData{iter}{N\_iter}
\SetKwData{Dist}{GP\_Dist}
\SetKwData{Prior}{Prior}
\SetKwData{Posterior}{Posterior}
\SetKwData{bv}{bestVal}
\SetKwData{cv}{curVal}
\SetKwData{np}{nextTheta}
\SetKwData{void}{void}
\SetKwFunction{ip}{InitPrior}
\SetKwFunction{SamplePoint}{SampleNextPoint}
\SetKwFunction{Update}{Update}
\SetKwFunction{Statistic}{Statistic}
\KwResult{optimal statistic $f_s(\theta_\text{opt})$}
\BlankLine
    \Dist $\leftarrow$ \ip\;
    \bv $\leftarrow$ \void\;
    \For{$i \leftarrow 1$ \KwTo \iter}{
        \np $\leftarrow$ \SamplePoint{\Dist}\;
        \tcp{calc. statistic with alg.~\ref{alg:sample_extreme_val_statistic}}
        \cv $\leftarrow$ \Statistic{\np}\;
        \If{\cv $>$ \bv}{\bv $\leftarrow$ \cv}
        \Dist $\leftarrow$ \Update{\Dist, \cv, \np}
    }

    \Return \bv\;
\caption{Bayesian Optimization of QAOA Extreme Value Statistics}\label{alg:bayes_opt}
\end{algorithm}

\subsection{Comparison to Random Sampling}

\begin{figure}[t!]
    \centering
    \subfloat[\label{app:fig:traces_randomized.}]{
        \includegraphics[width=0.48\textwidth]{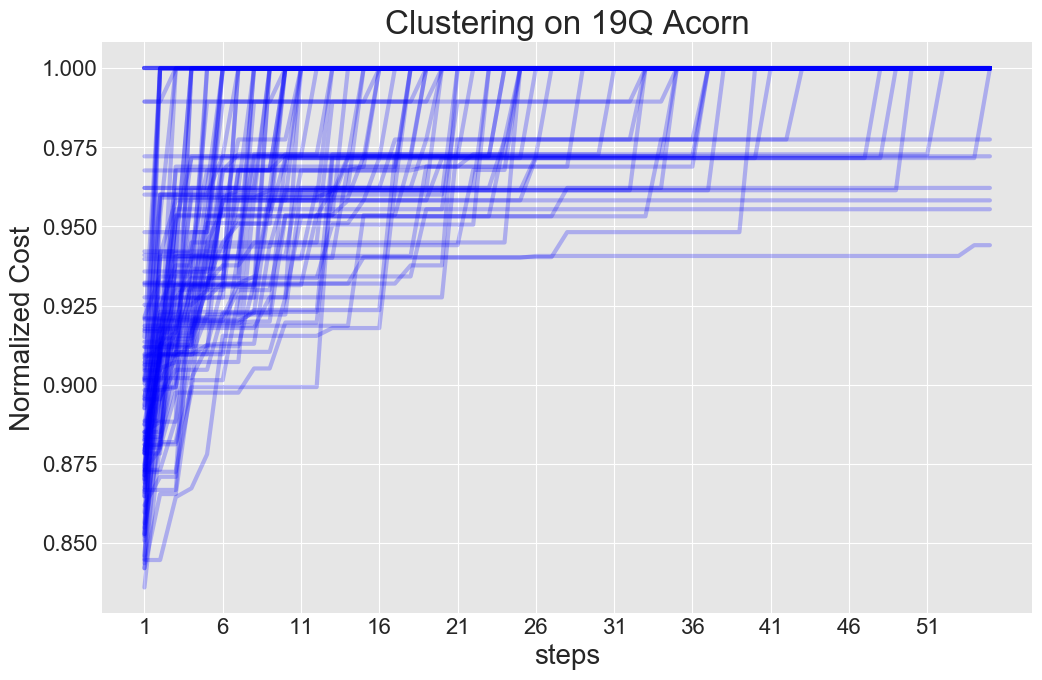}
    }
    ~
    \subfloat[\label{app:fig:ecdf}]{
        \includegraphics[width=.48\textwidth]{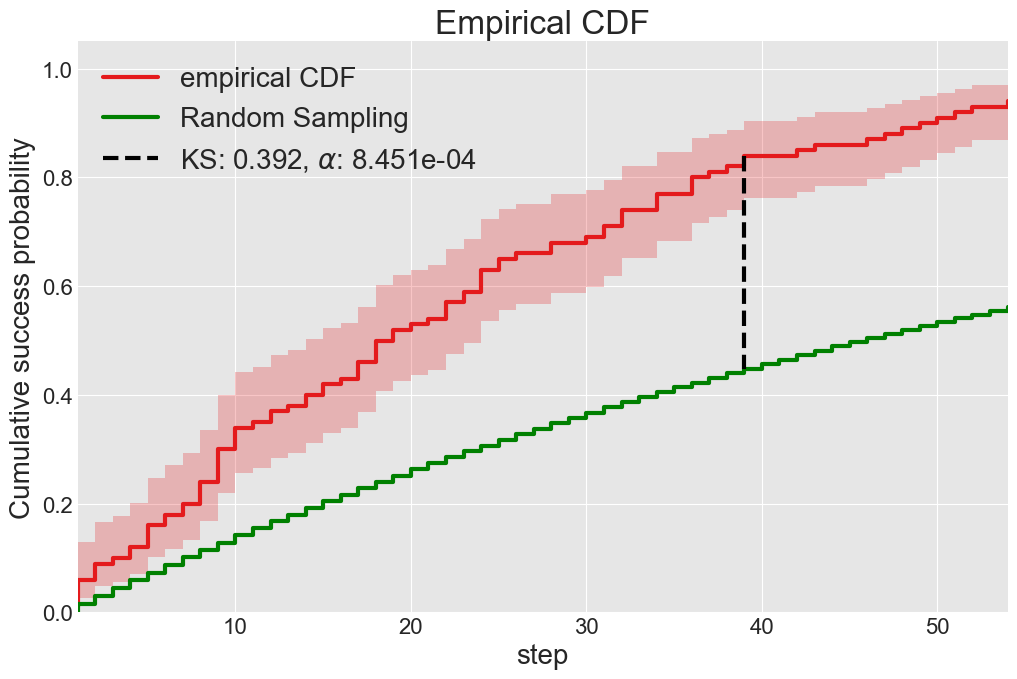}
    }
    \caption{(\textit{a}) Traces for the normalized \textsc{Maxcut} cost for 83 independent runs of the algorithm on the 19Q chip for a fixed random problem instances of Fig.~\ref{fig:max-cut-instance}. Notice that most traces reach the optimal cost well before the cutoff at 55 steps. (\textit{b}) The performance of our implementation of the clustering algorithm (red) can be compared to the performance of an algorithm that simply draws cluster assignments at random (green). It is clear that our algorithm generates the optimal assignment much more quickly than it would be expected by chance: the $95\%$ confidence region for our empirical observations have very small overlap for the distribution given by random assignments, and the Kolmogorov-Smirnov statistic indicates we can reject the null hypothesis of random assignments at a level higher than $99.9\%$.}
\end{figure}

To demonstrate the applicability of the results beyond a single problem instance we ran the simulations on 5 randomly chosen problem instances over a fourteen hour window on 19Q architecture. We recorded the optimization traces (cf. Fig.~\ref{app:fig:traces_randomized.}) and calculated the empirical CDF (eCDF) for the time-to-optimum, i.e. the number of steps before the optimizer reached the optimal value, as seen in Figs.~\ref{fig:ecdf-comparisson}, \ref{app:fig:ecdf}. Note that we can estimate the optimal value easily for the problem at hand. We compared the empirical CDF (eCDF) to the CDF of a random sampling procedure that follows a Bernoulli distribution $\mathcal{B}(N, p)$ with a success probability $p=2 / 2^{19}$ and $N=N_\text{steps} N_\text{shots}$ samples. The additional factor $2$ in the success probability is due to the inversion symmetry of the solution, i.e. there are two equivalent solutions which minimize the cost and are related to each other by simply inverting each bit-assignment. The CDF for the Bernoulli random variable can then be easily written as:

\begin{align}
    P(\text{success after $k$ steps}) = 1 - \left(1-p\right)^{k * N_\text{shots}}
    \label{app:eq:RandomSamplingCDF}
\end{align}

To compare the eCDF (red curve in Fig.~\ref{app:fig:ecdf}) to the random sampling CDF (green curve) we calculate the Kolmogorov-Smirnov statistic between two eCDFs

\begin{align}
    KS_{n,m} = \sup_{x}|F_{1,n}(x) - F_{2,m}(x)|
\end{align}
where $F_{1,n}$ is the first eCDF with $n$ points and $F_{2,m}$ is the second one with $m$ points. Given the eCDFs of Fig.~\ref{app:fig:ecdf} we find $KS_{23,55} \approx 0.392$. We can calculate the significance level $\alpha$ by inverting the prescription for rejection of the Null-Hypothesis $H_0$, i.e. the two eCDFs result from the same underlying distribution function:

\begin{align}
    KS_{n,m} \geq c(\alpha)\sqrt{\frac{n+m}{nm}}
\end{align}
where $c(\alpha) = \sqrt{-0.5 \log(\alpha/2)}$. Plugging in the empirical KS statistic we find that $H_0$ can be rejected with a probability $p=1-\alpha$ with $\alpha=8.451\cdot 10^{-4}$. We also calculated the KS statistic for the curves in the main body of the text summarized in Table.~\ref{app:table:ks_stats}

\begin{table}[t]
\caption{\label{app:table:ks_stats} Kolmogorov-Smirnov statistics and significance values for the CDF shown in Fig.~\ref{fig:ecdf-comparisson}. All values are calculated with respect to the exact random sampling CDF}
\begin{tabular}{r r r}
\hline\hline
 & & \\[-8pt]
eCDF  & KS & $\alpha$\\[3pt]
\hline \\[-9pt]
empirical random bitstring sampling (Fig.~\ref{fig:ecdf-comparisson}) & $0.077$ & $1.559$ \\[3pt]
Rigetti-QVM (Fig.~\ref{fig:ecdf-comparisson}) & $0.838$ & $1.273\cdot 10^{-7}$ \\[3pt]
19Q single instance (Fig.~\ref{fig:ecdf-comparisson}) & $0.339$ & $1.339 \cdot 10^{-2}$\\[3pt]
19Q randomized instances (Fig.~\ref{app:fig:ecdf}) & $0.392$ & $8.4451\cdot 10^{-4}$\\[3pt]
\hline\hline\\
\end{tabular}
\end{table}
\subsection{Example of a clustering application for the 19Q architecture}

\begin{figure}[t]
    \centering
    \includegraphics[width=0.75\textwidth]{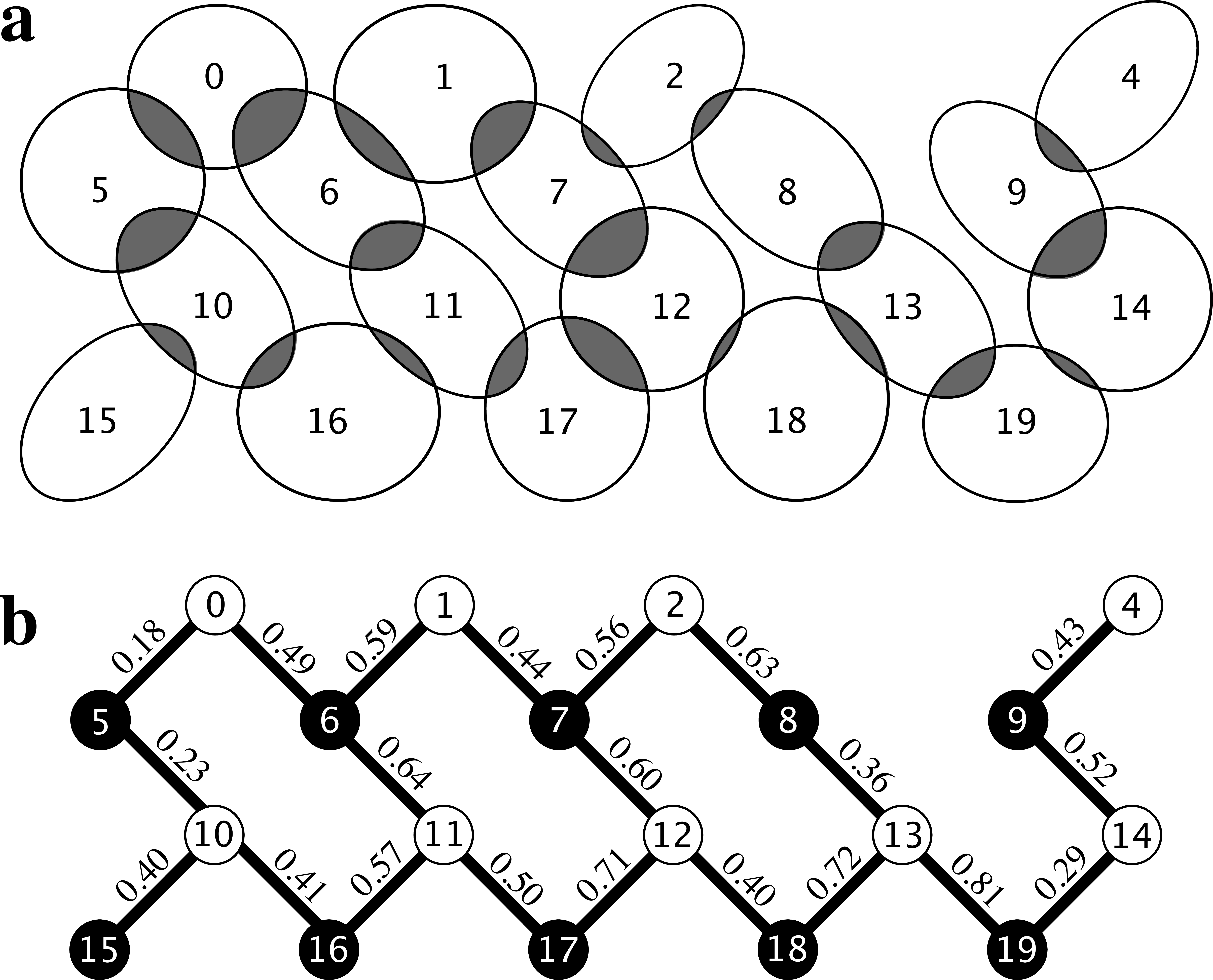}
    \caption{Pictorial representation of a set of compact distributions in $\mathbb{R}^2$. Calculating the mutual Bhattacharyya coefficient between the distributions leads to a graph that that maps to the Rigetti 19Q architecture with low overhead. Applying the proposed clustering algorithm will find assignments for sets of the least overlapping distributions. For more details see the body of the appendix.}
    \label{app:fig:distribution_overlap}
\end{figure}

To motivate the clustering application we look at Fig.~\ref{app:fig:distribution_overlap}. It shows 19 distribution with compact support in $\mathbb{R}^2$ with finite overlaps. We can define a similarity metric between these distributions using the Bhattacharyya coefficient
~\cite{Bhattacharyya1943, Chung1989, Coleman1979}
\begin{align}
    b: \mathcal{D}_{\mathbb{R}^2} \times \mathcal{D}_{\mathbb{R}^2} \longrightarrow \mathbb{R}, \quad b(p, q) = \int\sqrt{ p(x)q(x)} \textrm{d}^2x
\end{align}
where we use $\mathcal{D}_{\mathbb{R}^2}$ to denote the space of compact-support distributions over $\mathbb{R}^2$. Since the Bhattacharyya coefficient is not a typical distance metric --it does not fulfill the triangle inequality--, we use the following procedure: Using the example of Fig.~\ref{app:fig:distribution_overlap} we can identify the individual distributions with a given qubit and calculate the overlap metric with all other distributions. With this we can construct a graph $G=(V,E)$ where the vertices correspond to identifiers of the individual distributions and the edges to overlaps in these distributions where the weights are given by the Bhattacharyya coefficient. In the case of Fig.~\ref{app:fig:distribution_overlap} this will lead to a graph that has a low overhead when mapped to the Rigetti 19Q connectivity and enables us to run the clustering algorithm on the quantum hardware. To make this translation we need to remove the self-similarity between the distributions, corresponding to self-cycles in the graph. It should be noted that clustering in this context means to identify a set of distribution that are \emph{as dissimilar as possible}, i.e. have as little overlap as possible.

\subsection{Detailed device parameters}

\begin{table}[t]
\caption{\label{table:1} $\textbf{ Rigetti 19Q performance
parameters}$  | All of the parameters listed in this table have been measured at
base temperature $T\approx \unit[10]{mK}$. The reported $T_1$'s and $T^*_2$'s
are averaged values over 10 measurements acquired at $\omega^\textrm{max}_{01}$.
The errors indicate the standard deviation of the averaged value. Note that
these estimates fluctuate in time due to multiple factors.
}
\begin{tabular}{r r r r r r r r r r r }
\hline\hline
 & & \\[-8pt]
& $\;\omega^\textrm{max}_\textrm{r}/2\pi$ & $\quad\omega^\textrm{max}_\textrm{01}/2\pi$ & $\quad\eta/2\pi $ & $\quad T_1$ & $\quad T^*_2$ & $\qquad\mathcal{F}_\textrm{1q}$ & $\qquad\mathcal{F}_\textrm{RO}$ \\[2pt]
\hline
 & & \\[-6pt]
 & $\unit[]{MHz}$ &$\unit[]{MHz}$&$\unit[]{MHz}$  & $\unit[]{\mu s}$ &$\unit[]{\mu s}$&$\unit[]{ }$ & $\unit[]{ }$\\[3pt]
\hline \\[-9pt]
0  & 5592 & 4386 & -208 & $\quad\textbf{15.2} \pm 2.5$ & $\qquad\textbf{7.2} \pm 0.7$& $0.9815$ & $0.938$\\[3pt]
1  & 5703 & 4292 & -210 & $\textbf{17.6} \pm 1.7$& $\textbf{7.7} \pm 1.4$ & $0.9907$ & $0.958$\\[3pt]
2  & 5599 & 4221 & -142 & $\textbf{18.2} \pm 1.1$& $\textbf{10.8} \pm 0.6$& 0.9813 & 0.970\\[3pt]
3  & 5708 & 3829 & -224 & $\textbf{31.0} \pm 2.6$& $\textbf{16.8} \pm 0.8$& 0.9908 & 0.886\\[3pt]
4  & 5633 & 4372 & -220 & $\textbf{23.0} \pm 0.5$& $\textbf{5.2} \pm 0.2$& 0.9887 & 0.953\\[3pt]
5   & 5178 & 3690 & -224 & $\textbf{22.2} \pm 2.1$& $\textbf{11.1} \pm 1.0$& 0.9645 & 0.965\\[3pt]
6   & 5356 & 3809 & -208 & $\textbf{26.8} \pm 2.5$& $\textbf{26.8} \pm 2.5$& 0.9905 & 0.840\\[3pt]
7   & 5164 & 3531 & -216 & $\textbf{29.4} \pm 3.8$& $\textbf{13.0} \pm 1.2$& 0.9916 & 0.925\\[3pt]
8   & 5367 & 3707 & -208 & $\textbf{24.5} \pm 2.8$& $\textbf{13.8} \pm 0.4$& 0.9869 & 0.947\\[3pt]
9   & 5201 & 3690 & -214 & $\textbf{20.8} \pm 6.2$& $\textbf{11.1} \pm 0.7$& 0.9934 & 0.927\\[3pt]
10 & 5801 & 4595 & -194 & $\textbf{17.1} \pm 1.2$& $\textbf{10.6} \pm 0.5$& 0.9916 & 0.942\\[3pt]
11 & 5511 & 4275 & -204 & $\textbf{16.9}  \pm 2.0$& $\textbf{4.9} \pm 1.0$& 0.9901 & 0.900\\[3pt]
12 & 5825 & 4600 & -194 & $\textbf{8.2} \pm 0.9$& $\textbf{10.9} \pm 1.4$& 0.9902 & 0.942\\[3pt]
13 & 5523 & 4434 & -196 & $\textbf{18.7} \pm 2.0$& $\textbf{12.7} \pm 0.4$& 0.9933 & 0.921\\[3pt]
14 & 5848 & 4552 & -204 & $\textbf{13.9} \pm 2.2$& $\textbf{9.4} \pm 0.7$& 0.9916 & 0.947\\[3pt]
15  & 5093 & 3733 & -230 & $\textbf{20.8} \pm 3.1$& $\textbf{7.3} \pm 0.4$& 0.9852 & 0.970\\[3pt]
16  & 5298 & 3854 & -218 & $\textbf{16.7} \pm 1.2$& $\textbf{7.5} \pm 0.5$& 0.9906 & 0.948\\[3pt]
17  & 5097 & 3574 & -226 & $\textbf{24.0} \pm 4.2$& $\textbf{8.4} \pm 0.4$& 0.9895 & 0.921\\[3pt]
18  & 5301 & 3877 & -216 & $\textbf{16.9} \pm 2.9$& $\textbf{12.9} \pm 1.3$& 0.9496 & 0.930\\[3pt]
19  & 5108 & 3574 & -228 & $\textbf{24.7} \pm 2.8$& $\textbf{9.8} \pm 0.8$& 0.9942 & 0.930\\[3pt]
\hline\hline\\
\end{tabular}
\end{table}

\begin{table}[t]
\caption{\label{table:2} $\textbf{Rigetti 19Q two-qubit gate parameters and
performance}$  | These parameters refer to two-qubit interactions of Rigetti
19Q. Qubit 3 is not tunable and for this reason parameters related to the pairs
$3-8$, $3-9$ are not included.
}
\begin{tabular}{r r r r r r r r r r r }
\hline\hline
 & & \\[-8pt]
&$A_{0}$&$\quad f_\textrm{m}$& $\quad t_\textrm{CZ}$ & $\quad\mathcal{F}_\textrm{2q}\;$ & \\[2pt]
\hline
 & & \\[-6pt]
 &$\Phi/\Phi_0$& $\quad\unit[]{MHz}$&$\quad\unit[]{ns}$ & $\quad$& $\quad$&\\[3pt]
\hline \\[-9pt]
$0-5\;$   &0.27&94.5& 168 & $0.936\;$ & \\[3pt]
$0-6 \;$  &0.36&123.9& 197 & $0.889\;$ &\\[3pt]
$1-6 \;$  &0.37&137.1& 173 & $0.888\;$ &\\[3pt]
$1-7 \;$  &0.59&137.9 & 179 & $0.919\;$ &\\[3pt]
$2-7\;$   &0.62&87.4& 160 & $0.817\;$ &\\[3pt]
$2-8\;$   &0.23&55.6& 189 & $0.906\;$ &\\[3pt]
$4-9\;$   &0.43 & 183.6 & 122 & $0.854\;$ &\\[3pt]
$5-10\;$  &0.60&152.9& 145 & $0.870\;$ &\\[3pt]
$6-11\;$  &0.38&142.4& 180 & $0.838 \;$&\\[3pt]
$7-12\;$  &0.60&241.9& 214 & $0.870\;$ &\\[3pt]
$8-13\;$  &0.40&152.0& 185 & $0.881\;$ &\\[3pt]
$9-14\;$  &0.62&130.8& 139 & $0.872\;$ &\\[3pt]
$10-15\;$ &0.53&142.1& 154 & $0.854\;$ &\\[3pt]
$10-16\;$ &0.43&170.3& 180 & $0.838\;$ &\\[3pt]
$11-16\;$ &0.38&160.6& 155 & $0.891\;$ &\\[3pt]
$11-17\;$ &0.29&85.7& 207 & $0.844\;$ &\\[3pt]
$12-17\;$ &0.36&177.1& 184 & $0.876\;$ &\\[3pt]
$12-18\;$ &0.28&113.9& 203 & $0.886\;$ &\\[3pt]
$13-18\;$ &0.24&66.2& 152 & $0.936\;$ &\\[3pt]
$13-19\;$ &0.62&109.6& 181 & $0.921\;$ &\\[3pt]
$14-19\;$ &0.59&188.1& 142 & $0.797\;$ & \\[3pt]
\hline\hline
% & & \\[-10pt]
%\hline\hline\\
\end{tabular}
\end{table}

Tables~\ref{table:1} and~\ref{table:2} summarize the main performance parameters
of Rigetti 19Q. Single-qubit gate fidelities are estimated with standard
randomized benchmarking protocols~\cite{Magesan2011} with 25 random Clifford gate
sequences of lengths $l \in \{2,4,8,16,32,64,128\}$. Readout fidelity is given by the assignment fidelity
$\mathcal{F}_\textrm{RO} = [p(0|0)+p(1|1)]/2$, where $p(b|a)$ is the probability
of measuring the qubit in state $b$ when prepared in state $a$. Two-qubit gate
fidelities are estimated with quantum process tomography~\cite{Nielsen2002}
with preparation and measurement rotations $\{\mathbb{I}, R_x(\pi/2),
R_y(\pi/2), R_x(\pi)\}$. The reported process fidelity $\mathcal{F}_\textrm{2q}$
indicates the average fidelity between the ideal process and the measured
process imposing complete positivity and trace preservation
constraints. We further averaged over the extracted $\mathcal{F}_\text{2q}$ from
four separate tomography experiments. Qubit-qubit coupling strengths are
extracted from Ramsey experiments with and without $\pi$-pulses on neighboring
qubits.

\end{document}